\documentclass[showpacs,aps,prl,twocolumn,floatfix,%
superscriptaddress]{revtex4}
\usepackage{graphicx}
\usepackage{bm}
\usepackage{amsmath}
\begin{document}
\title{Berry phase in Magnetic Superconductors}
\author{Shuichi Murakami}
\email[Email: ]{murakami@appi.t.u-tokyo.ac.jp}
\affiliation{Department of Applied Physics, University of Tokyo,
Bunkyo-ku, Tokyo 113-8656, Japan}
\author{Naoto Nagaosa}
\affiliation{Department of Applied Physics, University of Tokyo,
Bunkyo-ku, Tokyo 113-8656, Japan}
\affiliation{CERC, AIST Tsukuba Central 4, Tsukuba 305-8562, Japan}
\date{\today}
\begin{abstract}
In magnetic systems, electronic bands often acquire nontrivial 
topological structure characterized by gauge flux distribution 
in momentum($\mathbf{k}$)-space. 
It sometimes follows that the phase of the wavefunctions
cannot be defined uniquely over the whole Brillouin zone.
In this Letter we develop a theory of superconductivity
in the presence of this gauge flux both in two- and 
three-dimensional systems. It is found that the superconducting
gap has ``nodes" as a function of 
$\mathbf{k}$ where the Fermi surface is penetrated by a gauge string.
\end{abstract}
\maketitle
\renewcommand{\labelenumi}{(\roman{enumi})}

Geometric phases and topology of 
wavefunctions have been the subject of intensive studies \cite{berry,review}. 
The quantized Hall effect (QHE) in a 2D electron system under
a strong magnetic field can be also interpreted in terms of the 
topological integer called Chern number \cite{tknn}.
If the system has a gap, the Hall conductivity is proportional to the Chern 
number and is quantized. This nontrivial topology of 
Bloch waves is not specific to the QHE, but is common and
universal in systems with broken time-reversal symmetry.
Namely, in magnetic materials with the spin-orbit interaction 
and/or tilting of spin configuration, 
there occur many band-crossing points
which are locally described as Weyl fermions and act as 
(anti)monopoles in the momentum-space 
\cite{ohgushi,onoda,shindou,taguchi}.
The existence of (anti)monopoles means that the Bloch wavefunction 
cannot be defined in a single gauge choice \cite{sakurai}. 
The anomalous Hall effect in ferromagnets represents
this topological property of the Bloch wavefunctions \cite{ohgushi,taguchi}.

On the other hand, coexistence of magnetic ordering and
superconductivity (SC) is found in many materials 
and is a recent significant issue \cite{super}. 
Although detailed analysis of each material is
still missing in most of the cases, 
it is highly desirable to establish the general
feature of the SC in systems with broken time-reversal symmetry.

\begin{figure}[h]
\includegraphics[scale=0.45]{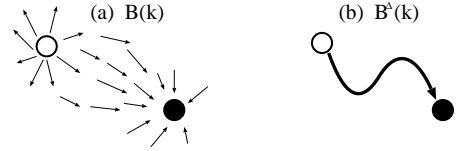}
\caption{Distribution of (a) $\mathbf{B}(\mathbf{k})$ and (b)
$\mathbf{B}^{\Delta}(\mathbf{k})$.
The open and the solid circles represent the monopoles and the antimonopoles,
respectively. 
}
\label{BB}
\end{figure}

In the present Letter we shall study the SC
made out of the Bloch states with nontrivial topology.
The quantity of central importance is the gauge flux defined 
in the momentum space, which is generated by the vector potential 
$
\mathbf{A}_{(n)} (\mathbf{k}) = - i \langle\mathbf{k} n | \nabla_{\mathbf{k}}
| \mathbf{k} n \rangle
$,
where $ | \mathbf{k} n \rangle $ is the Bloch wavefunction 
of the $n$-th band. 
Band indices are represented by subcripts with parenthesis.
This $ \mathbf{A}_{(n)} (\mathbf{k})$ represents an 
overlap of the two wavefunctions separated infinitestimally 
in the $\mathbf{k}$-space, i.e. the Berry phase connection.
The gauge flux is defined as 
$\mathbf{B}_{(n)}( \mathbf{k}) = \nabla_{\mathbf{k}} \times \mathbf{A}_{(n)} (\mathbf{k})$.
It is worth noting that the monopole density
$\rho_{(n)}(\mathbf{k})=
\frac{1}{2\pi}\nabla_{\mathbf{k}} \cdot  \mathbf{B}_{(n)} (\mathbf{k})$
is nonzero, and is given by 
$\rho_{(n)} (\mathbf{k})=\sum_{l}q_{l n}\delta(\mathbf{k}-\mathbf{k}_{l n})$, where
$q_{ln}$ is an integer called the
strength of the monopole \cite{berry}.
The monopoles are located at 
``diabolical points'' \cite{berry,volovik}, where
bands touch each other. 
The monopole and antimonopole act as the source and the sink of the 
gauge flux $\mathbf{B}(\mathbf{k})$, respectively, as in
Fig.~\ref{BB}(a).
In general, the up- and down-spins have different band structures;
therefore, $\mathbf{A}$, $\mathbf{B}$, and $\rho$ are spin-dependent, and 
we shall write as $\mathbf{A}_{\alpha}$, $\mathbf{B}_{\alpha}$ and 
$\rho_{\alpha}$, where $\alpha$ is a spin index.

The question we address below is an effect of this 
gauge flux $\mathbf{B}(\mathbf{k})$ on the SC properties.
In order to characterize topological properties of the gap function 
$\Delta(\mathbf{k})$, we define
a vector potential  
$\mathbf{A}^{\Delta}(\mathbf{k})=-\nabla_{\mathbf{k}}\text{Im}\ln\Delta(\mathbf{k})=
-\nabla_{\mathbf{k}}\text{arg}\Delta(\mathbf{k})$, 
a flux density $\mathbf{B}^{\Delta}=\nabla_{\mathbf{k}}\times
\mathbf{A}^{\Delta}(\mathbf{k})$, and 
a monopole 
density 
$\rho^{\Delta}
(\mathbf{k})=\frac{1}{2\pi}\nabla_{\mathbf{k}}
\cdot\mathbf{B}^{\Delta}(\mathbf{k})$.
For the moment, we assume that $\Delta(\mathbf{k})$ is nonvanishing
in almost all over the BZ \cite{noteDelta}; 
we shall discuss later what happens without this assumption.
As explained later, distribution 
of $\mathbf{B}^{\Delta}(\mathbf{k})$ is confined into strings as in 
Fig.~\ref{BB}(b), due to bosonic nature of $\Delta(\mathbf{k})$.
The main result of the present Letter is that in 3D 
topological structures of the gap function $\Delta(\mathbf{k})$ is 
solely determined from that of the wavefunction: 
$\rho^{\Delta}_{\alpha\beta}(\mathbf{k})
=\rho_{\alpha}(\mathbf{k})-\rho_{\beta}(-\mathbf{k})$.
It reflects the fact that $\Delta_{\alpha\beta}(\mathbf{k})$ represents a pairing 
between $(\mathbf{k},\alpha)$ and $(-\mathbf{k},\beta)$ electrons.
Thus if the band structure 
in the normal 
state is known, we can immediately calculate the monopole density 
$\rho^{\Delta}_{\alpha\beta}(\mathbf{k})$, irrespective of 
details of the attractive potential.
The monopole density $\rho^{\Delta}_{\alpha\beta}(\mathbf{k})$ 
tells us about zeros and phases of the gap function.

Henceforth we focus on the SC where only one component of the 
gap function is nonzero:
(i) singlet SC, 
(ii) triplet SC with $\Delta_{\uparrow\downarrow}\neq 0$, 
$\Delta_{\uparrow\uparrow}= 
\Delta_{\downarrow\downarrow}=0$, and (iii) triplet SC with
$\Delta_{\uparrow\uparrow}\neq 0$, 
$\Delta_{\uparrow\downarrow}= 
\Delta_{\downarrow\downarrow}=0$ \cite{note-triplet}.
The case (iii) is appropriate for half-metallic magnetic SC. 
In the singlet (triplet) SC, we have 
$\Delta(\mathbf{k})=\Delta(-\mathbf{k})$
($\Delta(\mathbf{k})=-\Delta(-\mathbf{k})$), 
implying 
$\mathbf{B}^{\Delta}
(\mathbf{k})=\mathbf{B}^{\Delta}(-\mathbf{k})$, and 
$\rho^{\Delta}(\mathbf{k})=-\rho^{\Delta}
(-\mathbf{k})$ \cite{note-delta}. We note that 
{\it the time-reversal symmetry  breaking is a necessary condition for
nontrivial topology of the gap function}.
In (iii), broken time-reversal symmetry is 
assumed from the 
outset. We have $\rho_{\uparrow\uparrow}^{\Delta}(\mathbf{k})
=\rho_{\uparrow}(\mathbf{k})-\rho_{\uparrow}(-\mathbf{k})$.
Thus, inversion symmetry must be broken to have 
nonzero $\rho^{\Delta}_{\uparrow \uparrow}$. 
In (i) and in (ii)
we have $\rho^{\Delta}_{\uparrow\downarrow}
(\mathbf{k})=\rho_{\uparrow}(\mathbf{k})-\rho_{\downarrow}(-\mathbf{k})$.
Hence in nonmagnetic SC, 
$\rho^{\Delta}_{\uparrow\downarrow}(\mathbf{k})$ vanishes.
In magnetic SC, including both ferromagnets and antiferromagnets, 
$\rho_{\uparrow\downarrow}^{\Delta}(\mathbf{k})$ is
nonzero.

\begin{figure}[h]
\includegraphics[scale=0.7]{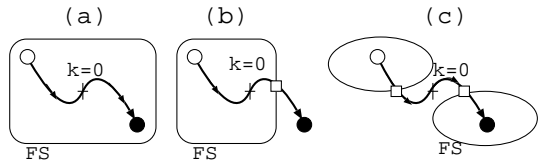}
\caption{Monopoles (open circles), antimonopoles (solid circles) for 
$\Delta(\mathbf{k})$, and 
the Fermi surface. The open squares represent point nodes 
on the Fermi surface.}
\label{fig:FS}
\end{figure}
Let us consider the distribution of $\mathbf{B}^{\Delta}(\mathbf{k})$.
$\mathbf{B}^{\Delta}(\mathbf{k})=
-\nabla_{\mathbf{k}}\times\nabla_{\mathbf{k}} \text{arg}\Delta(\mathbf{k})$
is nonzero only 
when $\Delta=0$, i.e. $\text{Re}\Delta=\text{Im}\Delta=0$, 
which determines a curve in the
3D BZ in general (Fig.~\ref{BB} (b)). 
Distribution of $\mathbf{B}^{\Delta}
(\mathbf{k})$ is like a Dirac string, starting from a monopole
and terminating at an antimonopole.
In contrast, $\mathbf{B}(\mathbf{k})$ does not share this feature;
distribution of $\mathbf{B}(\mathbf{k})$ spreads over the BZ (Fig.~\ref{BB}
(a)).
Now we discuss the relation among the Fermi surface (FS), 
the (anti)monopoles and the 
Dirac string. For any closed surface $S$ in the BZ, the total flux penetrating 
it is quantized. We take the FS as $S$ 
for gapless systems.
When the total flux penetrating one of the FS's 
is nonzero, the gap function cannot be expressed 
as continuous. It is quite contrary to our conventional view
\cite{note-2}. 
Because $\rho^{\Delta}(\mathbf{k})=-\rho^{\Delta}(-\mathbf{k})$, 
a FS symmetric with respect to $\mathbf{k}=0$ 
encloses zero monopoles in total, and $\Delta(\mathbf{k})$ can 
be continuous on this FS (Fig.~\ref{fig:FS}(a)). 
There are, however, two cases where the Dirac strings penetrate
the FS's. 
One is the case with broken inversion as well as time-reversal symmetry.
Then, the FS is not symmetric with respect to $\mathbf{k}=0$, and can 
enclose monopoles with nonvanishing strength (Fig.~\ref{fig:FS}
(b)). The other is a pair of FS's symmetric with respect to 
$\mathbf{k}=0$. It is possible that one FS encompasses
a monopole and the other encompasses an antimonopole (Fig.~\ref{fig:FS}(c)).
For example, if one of the FS encloses a monopole of unit strength, 
a Dirac string starting from this monopole necessarily penetrates the 
FS.  An intersection of this curve with the FS is nothing but a point node. 
This string can intersect the FS more than 
once, and the number of point nodes on this FS should be odd.
The nonvanishing total flux implies that $\Delta(\mathbf{k})$ 
cannot be a single continuous function on each FS. 
The FS should be divided into regions, in each of which $\Delta$ is 
continuous.

Here we remark on ``conventional'' nodes in anisotropic SC.
Line nodes appear when
$\text{Re}\Delta$ and $\text{Im}\Delta$
have a common real factor $f(\mathbf{k})$ as
$\Delta(\mathbf{k})=f(\mathbf{k})[\text{Re}g(\mathbf{k})+i\text{Im}g(\mathbf{k})]$. 
Zeros of $f(\mathbf{k})$ determine a surface,
whose intersection with the FS is a line node.
It does not affect $\mathbf{B}^{\Delta}(\mathbf{k})$, 
because $\nabla
\text{Im}\ln\Delta(\mathbf{k})=\nabla\text{Im}\ln g(\mathbf{k})$.
Hence, only point nodes are concluded from the gauge flux argument.
Difference of nonmagnetic SC from magnetic SC lies only in 
absence of magnetic monopoles; in nonmagnetic SC, Dirac strings 
necessarily form loops, which may cross the boundary of the BZ.
There are many experimental methods to get information on point nodes.
In particular, heat conductivity measurements 
tell us about the direction of the nodes \cite{izawa}.
When the positions of the nodes are known, care must be taken to 
assign the phase of $\Delta(\mathbf{k})$ for magnetic SC, 
because our theory asserts that $\Delta(\mathbf{k})$ need not be 
continuous on the FS.
Group-theoretical classifications of gap functions in \cite{su,vg} do
not apply to the SC with nontrivial topology, since
they assume that gap functions are continuous.
It would be interesting to classify all possible multivalued gap
functions. 

To see why topological structure of the wavefunction 
is inherited by the gap function, we note the following.
If the wavefunction is a continuous function of
$\mathbf{k}$, there is no monopole.
Thus, if there is a monopole, 
we should divide a surface surrounding it
into regions, in each of which the wavefunction
is continuous \cite{kohmoto}.
This resembles 
Dirac monopoles \cite{wy}, where the vector potential 
cannot be a single continuous function in the whole space.
This ``patch'' structure of the wavefunctions remains in the 
gap function via the BCS term in the Hamiltonian.

\begin{figure}[h]
\includegraphics[scale=0.42]{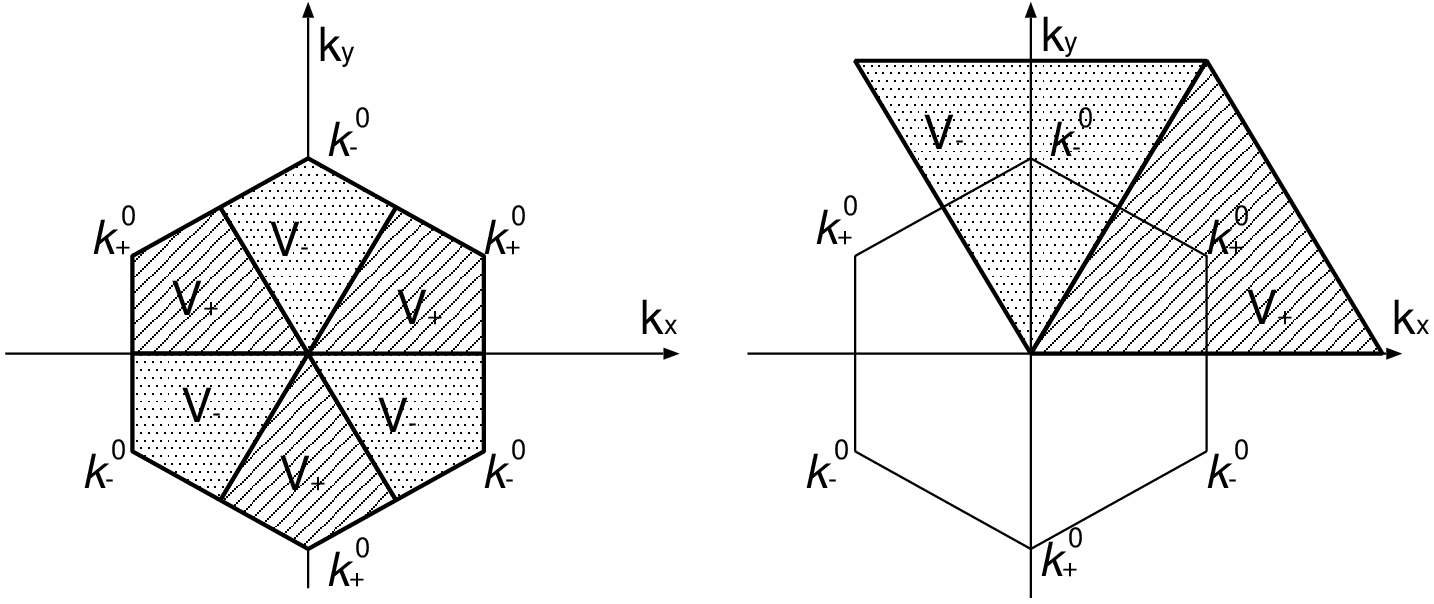}
\caption{Division of the Brillouin zone of the Haldane model into 
two regions $V_{+}$ and $V_{-}$.}
\label{BZ}
\end{figure}

To understand this mechanism in detail,
we start with a 2D model and generalize it to 3D.
Similar topological arguments as we have developed in 3D are 
also possible in 2D as well.
In 2D, we define a Chern number and 
a total vorticity of $\Delta(\mathbf{k})$ as 
$\text{Ch}=\frac{1}{2\pi}\int_{\text{BZ}} d^{2}\mathbf{k} B
(\mathbf{k})_{z}$ and
$\nu=\frac{1}{2\pi}\int_{\text{BZ}} d^{2}\mathbf{k} B^{\Delta}
(\mathbf{k})_{z}$, respectively.
We can show that the total vorticity $\nu_{\alpha\beta}$ of
the gap function $\Delta_{\alpha\beta}$ 
is the sum of the Chern numbers for spin $\alpha$ and
for spin $\beta$:  $\nu_{\alpha\beta}=\text{Ch}_{\alpha}+
\text{Ch}_{\beta}$ \cite{note-2D}.
We shall explain how this result comes out
in the 
singlet SC on the 2D model 
proposed by Haldane \cite{haldane} as an illustrative example. 
We take this model because it is the minimal model 
which exhibits 
nontrivial topological structure of wavefunctions
without a uniform magnetic field.
It is a tight-binding model on the honeycomb lattice.
Because there are two sublattices,
the Hamiltonian is written as a $2\times 2$ matrix, which 
is conveniently expressed with the Pauli matrices $\bm{\sigma}$ as
$H(\mathbf{k})=vI+\mathbf{w}\cdot\bm{\sigma}$, where 
$v=2t_{2}\cos\phi\sum_{j}\cos(\mathbf{k}\cdot\mathbf{b}_{j})$,
$w_{1}=t_{1}\sum_{j}\cos(\mathbf{k}\cdot\mathbf{a}_{j})$, 
$w_{2}=t_{1}\sum_{j}\sin(\mathbf{k}\cdot\mathbf{a}_{j})$, and
$w_{3}=M-2t_{2}\sin\phi\sum_{j}\sin(\mathbf{k}\cdot\mathbf{b}_{j})$,
where 
$\mathbf{a}_{1}=a(1,0)$, $\mathbf{a}_{2}=a(-\frac{1}{2},\frac{\sqrt{3}}{2})$,
$\mathbf{a}_{3}=-\mathbf{a}_{1}-\mathbf{a}_{2}$, 
$\mathbf{b}_{1}=\mathbf{a}_{2}-\mathbf{a}_{3}$,
$\mathbf{b}_{2}=\mathbf{a}_{3}-\mathbf{a}_{1}$,
$\mathbf{b}_{3}=\mathbf{a}_{1}-\mathbf{a}_{2}$. 
$t_{1}$, $t_{2}$, $M$, $\phi$ are
constants.
The Pauli matrices are not associated with spins,
but with the sublattice structure. 
We set $M=0$ and $0<\phi<\pi$.
Time-reversal symmetry is broken by a staggered flux, 
while the uniform field is absent.
This staggered flux is not necessarily external; staggered flux
can be realized by a spontaneous magnetic moment in the material itself,
through spin-chirality and/or spin-orbit coupling \cite{ohgushi,taguchi}.
Let us calculate the Chern number for the lower band.
By dividing the BZ into two regions $V_{\pm}$
as in Fig.~\ref{BZ}, 
where $\mathbf{k}_{\pm}^{0}=\pm\frac{2\pi}{a}(\frac{1}{3},\frac{1}{3\sqrt{3}})$ 
are two
inequivalent corners of the BZ,
we can write down the eigenvector
of the lower band as
$\mathbf{v}_{+}=\frac{1}{\sqrt{2|\mathbf{w}|(|\mathbf{w}|-w_{3})}}
\left(
\begin{array}{r}
w_{3}-|\mathbf{w}| \\
w_{1}+iw_{2}
\end{array}
\right)$, 
for $\mathbf{k}\in V_{+}$, and 
$\mathbf{v}_{-}=\frac{1}{\sqrt{2|\mathbf{w}|(|\mathbf{w}|+w_{3})}}
\left(
\begin{array}{r}
-w_{1}+iw_{2} \\
w_{3}+|\mathbf{w}|
\end{array}
\right)$, for $\mathbf{k}\in V_{-}$.
There remains some freedom in the division of the BZ.
Because $\mathbf{v}_{\pm}$ is ill-defined only at $\mathbf{k}_{\mp}^{0}$,
respectively,
we are free to 
deform this division  as long as 
$\mathbf{k}^{0}_{\mp}\not\in V_{\pm}$. This corresponds to 
the gauge degree of freedom. 
At  $\mathbf{k}\in V_{+}\cap V_{-}$, two choices of 
the wavefunction are different by a phase factor
$e^{i\phi(\mathbf{k})}\mathbf{v}_{+}=\mathbf{v}_{-}$, i.e.,
$\mathbf{A}_{+}-\mathbf{A}_{-}=-\nabla_{\mathbf{k}} \phi(\mathbf{k})$, 
where $
e^{i\phi(\mathbf{k})}=
\frac{w_{1}-iw_{2}}{\sqrt{w_{1}^{2}+w_{2}^{2}}}$.
Thus, we get 
$\text{Ch}=
\oint_{\partial V_{+}}(
\mathbf{A}_{+}-\mathbf{A}_{-})\cdot 
\frac{\text{d}\mathbf{k}}{2\pi}
=
-\oint_{\partial V_{+}}\frac{\text{d}\phi(\mathbf{k})}{2\pi}=-1$ \cite{haldane}.
The nonzero Chern number implies 
that the wavefunction cannot be written as a single function for the 
entire BZ. 
This also affects the definition of field operators $a_{(i)\mathbf{k}}$.
Second-quantized Hamiltonian is written as 
${\cal H}=\sum_{\mathbf{k},i,j}c_{\mathbf{k}j}^{\dagger}H_{ji}(\mathbf{k})c_{\mathbf{k}i}$.
Field operators $a_{(i)\mathbf{k}}$ are written as
$a_{(i)\mathbf{k}}
=\sum_{j}v^{\dagger}_{(i)j}(\mathbf{k})c_{j\mathbf{k}}
$.
Let $a_{\mathbf{k}\pm}$ denote the annihilation operators for the lower 
band when $\mathbf{k}\in V_{\pm}$.
Then, $e^{i\phi(\mathbf{k})}\mathbf{v}_{+}=\mathbf{v}_{-}$ yields
$a_{\mathbf{k}+}=e^{i\phi(\mathbf{k})}
a_{\mathbf{k}-}$ at $\mathbf{k}\in V_{+}\cap V_{-}$.

We now consider the singlet SC on the lower band.
In conventional formalism in the BCS theory of SC,
it is assumed that the field operators are 
continuous in the whole BZ;
it is no longer true in the present case. 
We assume that the Fermi energy crosses 
only the lower band.
The BCS pairing term in the Hamiltonian is 
\begin{equation}
\sum_{\mathbf{k}\in V_{+}}\Delta(\mathbf{k})_{+}
a_{\mathbf{k}+\uparrow}^{\dagger}
a_{-\mathbf{k}-\downarrow}^{\dagger}
+\sum_{\mathbf{k}\in V_{-}}\Delta(\mathbf{k})_{-}
a_{\mathbf{k}-\uparrow}^{\dagger}
a_{-\mathbf{k}+\downarrow}^{\dagger}.
\end{equation}
Singlet SC implies that $
\Delta(\mathbf{k})_{+}=\Delta(-\mathbf{k})_{-}$ for  $\mathbf{k}\in V_{+}$, i.e.
$\Delta$ is an even function of $\mathbf{k}$.
The guiding principle to study topological structure of 
$\Delta(\mathbf{k})$ is the gauge invariance, corresponding to the freedom
in the division of the BZ.
It is equivalent to require that the BCS term 
is continuous at the overlap of two regions; 
we get
$\Delta(\mathbf{k})_{+}=e^{2i\phi(\mathbf{k})}\Delta(\mathbf{k})_{-}$ for  
$\mathbf{k}\in V_{+}\cap V_{-}$.
The total vorticity $\nu$ is 
\begin{eqnarray}
&&\nu=
-\sum_{m=\pm}\oint_{\partial V_{m}}
\frac{\text{d}\text{arg}\Delta(\mathbf{k})_{m}}{2\pi}
=\frac{-1}{2\pi}\oint_{\partial V_{+}}\text{d}\text{arg}
\left(
\frac{\Delta_{+}}{\Delta_{-}}\right)
\nonumber \\
&&
=
-2\oint_{\partial V_{+}}\frac{\text{d}\phi(\mathbf{k})}{2\pi}=2\text{Ch}=-2.
\end{eqnarray}
Nonzero $\nu$ implies that 
$\Delta(\mathbf{k})$ can never be expressed as a single continuous  
function for the entire BZ. 
To see what the gap function looks like, we take an example of
an on-site attraction: 
${\cal H}_{U}=-U\sum_{i}n_{i\uparrow}n_{i\downarrow}$.
We assume that 
there are two hole pockets surrounding
$\mathbf{k}_{\pm}^{0}$ each.
The attractive potential $V(\mathbf{k},\mathbf{k}')$ for the lower band 
is factorized as 
a product of a function of $\mathbf{k}$ and that of $\mathbf{k}'$.
We obtain thereby
$
\Delta(\mathbf{k})_{\pm}=C
\frac{w_{1}(\mathbf{k})
\mp iw_{2}(\mathbf{k})}{|\mathbf{w}(\mathbf{k})|}\ \ \ (\mathbf{k}\in V_{\pm})$,
where $C$ is a complex constant.
This has two vortices at $\mathbf{k}=\mathbf{k}_{\pm}^{0}$, 
with vorticity $-1$ each.
The spin Hall conductivity $\sigma_{xy}^{s}$  \cite{spinHall}
can be calculated as $\sigma_{xy}^{s}=0$;
it is quantized when $\Delta(\mathbf{k})\neq 0$ on the FS,
and $\sigma_{xy}^{s}$ can be nonzero in general.

We can construct simple 3D models starting from the Haldane's model. 
For example, we set
$w_{1}=t_{1}\sum_{j}\cos (\mathbf{k}\cdot \mathbf{a}_{j})$, 
$w_{2}=t_{1}\sum_{j}\sin (\mathbf{k}\cdot \mathbf{a}_{j})$, 
$w_{3}=-2t_{2}\cos(k_{z}c)\sum_{j}\sin (\mathbf{k}\cdot \mathbf{b}_{j})$
in the  
Hamiltonian $H=\mathbf{w}\cdot\bm{\sigma}$, where
$c$ is the lattice constant
in the $c$-direction. 
This is a tight-binding 
model on a stacked honeycomb lattice with hopping along
$\pm\mathbf{a}_{j}$ (nearest-neighbor) and $\pm\mathbf{b}_{j}\pm (0,0,c)$;
the latter hopping acquires $\pi/2$ phase if it is
clockwise in the hexagonal plaquette.
This model has
$\rho=\sum_{\alpha,\beta=\pm}\beta
\delta(\mathbf{k}-\mathbf{k}_{\alpha}^{0}+\beta\frac{\pi}{2c}\hat{\mathbf{c}})
$, leading to $\rho^{\Delta}=
2\rho\neq 0$. In 
some range of parameters, some of the FS's encircle a monopole or antimonopole,
and the total flux is nonzero for them. 

So far we have dealt with noninteracting systems. One may wonder if the 
topological properties of $\Delta(\mathbf{k})$ discussed above are robust
against interactions. In general, robustness of physical 
phenomena originated from topology is guaranteed by an 
existence of integer topological numbers, which remain unchanged
under an adiabatic change of the Hamiltonian.
Although this adiabatic principle usually applies
to gapful systems,
it applies also to Fermi liquid with 
interactions.
In Fermi liquid,
bare electrons turn to quasiparticles which are 
well-defined near the FS. 
(Even with interactions, the FS is topologically stable 
in the momentum space. See Fig.~1 in \cite{volovik2}.)
The BCS theory is founded on this Fermi liquid theory,
and the gap function $\Delta(\mathbf{k})$ is defined 
in terms of these quasiparticles. 
The total strength of $\Delta(\mathbf{k})$-monopoles inside the FS is
a surface integral of a well-defined function $-\frac{1}{2\pi}
\nabla_{\mathbf{k}}\times
\nabla_{\mathbf{k}} \text{arg}
\Delta(\mathbf{k})$ over the 
FS, and it remains a well-defined integer 
even when we include interactions and when 
$\Delta(\mathbf{k})$ vanishes far from the FS.
When we start with 
noninteracting systems, and adiabatically switch on interactions, 
this integer remain the same for certain range of 
the strength of the interactions.
We note that also 
the total strength of monopoles for the wavefunction inside the FS
remains an integer in the presence of interactions.
Its generalized definition in the presence of interactions is 
$
\frac{1}{24\pi^{2}}\epsilon_{\mu\nu\lambda\gamma}\text{Tr}\int_{S}
dS^{\lambda}G\partial_{k_{\mu}}G^{-1}
G\partial_{k_{\nu}}G^{-1}
G\partial_{k_{\lambda}}G^{-1}
$
where $G(i\omega, \mathbf{k})$ is the one-particle 
Matsubara Green's function, and 
$S$ is the three-dimensional surface surrounding 
the FS in the four-dimensional frequency-momentum space
$(\omega,\mathbf{k})$ (see Eq.~(65) in \cite{volovik2}). 
It is a well-defined integer, and
topologically stable \cite{volovik2}.
After applying the Stokes' theorem, 
the integrand is finite only on the FS, and 
the integral is equal to the number of monopoles inside the FS.
In 3D,
Fermi liquid picture is valid even with interactions, and 
modification of Fermi velocity 
$v_{F}$ or of quasiparticle residue $Z$ does not change the number of
monopoles inside the FS. Thus the 
total monopole strength remains unchanged.
It changes only when the topology of the FS's
changes.

Let us discuss on applicability of our theory to real materials.
As mentioned above, we should search through magnetic SC with 
(a) equal-spin pairing without inversion
symmetry or  (b) opposite-spin pairing. The case (a) is 
likely in ferromagnetic SC, while (b) is likely in antiferromagnetic SC.
Band structure calculations have been concentrating on 
energy eigenvalues on high-symmetry directions.
Almost no attention has been paid to existence of band-touching 
points, or to phases of wavefunctions. Thus, we do not 
have at present
sufficient information to see which materials are candidates for
our theory. In order to consider how often nontrivial 
topological structure appears, we should instead resort to  
simplified models.
In \cite{onoda}, the authors consider a ferromagnetic 
tight-binding model on the 2D square lattice with $t_{2g}$ orbitals,
with physically reasonable values of parameters. 
The gauge flux $B(\mathbf{k})_{z}$ has sharp peaks,
where the bands approach each other in energy. 
If we go to the 3D by adding an extra dimension along $k_{z}$, 
the bands will touch each other at some points,
because band-touching points (diabolical points) 
has codimension three \cite{volovik,volovik2}. Actually in \cite{shindou},
the diabolical points, i.e. the (anti)monopoles, are
found in a model for an antiferromagnet.
These points are topologically stable.
Thus nontrivial topology of the wavefunctions and the gap 
functions, as discussed in the present Letter, is expected 
in many materials.

We can argue exciton condensation in the similar way. 
In 2D, total vorticity of the order parameter 
is equal to the difference of Chern numbers of the bands involved. 
In 3D, the monopole density of the order parameter
is the difference of those for wavefunctions of the two bands.
  
In conclusion, we considered the SC in a band with nontrivial topology.
It is formulated in terms of a gauge flux distribution in the $\mathbf{k}$-space.
Although the gap function $\Delta(\mathbf{k})$
depends on a detail of attractive 
potential, topological structure of $\Delta(\mathbf{k})$ is 
determined solely from that of the normal-state wavefunction.
The effect of disorder is an interesting issue, and is left for 
future studies.

\begin{acknowledgments}
The authors thank helpful discussion with R.~Shindou 
and M.~Sigrist. We acknowledge support by 
Grant-in-Aids
from the Ministry of Education, Culture, Sports, Science and Technology
of Japan. 
\end{acknowledgments}


\begin{thebibliography}{99}
\bibitem{note-triplet}
In general triplet SC, we consider $\text{det}
\hat{\Delta}=\Delta_{\uparrow\uparrow}
\Delta_{\downarrow\downarrow}
-\Delta_{\uparrow\downarrow}\Delta_{\downarrow\uparrow}$, which 
is a product of two gaps.
It has a monopole density $\rho_{\uparrow}(\mathbf{k})+
\rho_{\downarrow}(\mathbf{k})
-\rho_{\uparrow}(-\mathbf{k})-\rho_{\downarrow}(-\mathbf{k})$.
\bibitem{noteDelta}
It means that there is no 3D region with 
$\Delta(\mathbf{k})=0$.
\bibitem{note-delta}
In the pairing with opposite spins $\Delta_{\uparrow\downarrow}\neq 0$, 
this gives a 
constraint $\rho_{\uparrow}(\mathbf{k})+\rho_{\uparrow}(-\mathbf{k})
=\rho_{\downarrow}(\mathbf{k})+\rho_{\downarrow}(-\mathbf{k})$. Therefore, 
the pairing with opposite spins is prohibited 
when the normal-state wavefunction does not satisfy this relation.
\bibitem{note-2}
As for the freedom of local gauge transformation 
$\Delta(\mathbf{k})\rightarrow \Delta(\mathbf{k})e^{i\varphi(\mathbf{k})}$,
it is natural to choose the gauge so that   $\Delta(\mathbf{k})$
is continuous, if possible.
What we have shown is that in some cases $\Delta$ cannot be
continuous in any gauge choice.
\end{thebibliography}
\end{document}